\documentclass[conference]{IEEEtran}
\IEEEoverridecommandlockouts
\usepackage{cite}
\usepackage{amsmath,amssymb,amsfonts}
\usepackage{algorithmic}
\usepackage{graphicx}
\usepackage{textcomp}

\usepackage{xcolor}
\def\BibTeX{{\rm B\kern-.05em{\sc i\kern-.025em b}\kern-.08em
    T\kern-.1667em\lower.7ex\hbox{E}\kern-.125emX}}
\begin{document}

\title{FedMon: Federated eBPF Monitoring for Distributed Anomaly Detection in Multi-Cluster Cloud Environments\\
}

\author{\IEEEauthorblockN{Sehar Zehra}
\IEEEauthorblockA{\textit{Department of Computer Science} \\
\textit{FAST National University of Computer}\\
{\& Emerging Sciences}\\
Karachi, Pakistan \\
0009-0007-7595-1221}
\and
\IEEEauthorblockN{Hassan Jamil Syed}
\IEEEauthorblockA{\textit{Asia Pacific University of } \\
{ Technology\& Innovation (APU)}\\
Kuala Lumpur, Malaysia \\
 0000-0002-1834-1810}
\and
\IEEEauthorblockN{ Ummay Faseeha}
\IEEEauthorblockA{\textit{Department of Computer Science} \\
\textit{FAST National University of Computer}\\
{\& Emerging Sciences}\\
Karachi, Pakistan \\
0009-0000-5276-1504}
\and

}

\maketitle

\begin{abstract}
Kubernetes multi-cluster deployments demand scalable and privacy-preserving anomaly detection. Existing eBPF-based monitors provide low-overhead system and network visibility but are limited to single clusters, while centralized approaches incur bandwidth, privacy, and heterogeneity challenges. We propose FedMon, a federated eBPF framework that unifies kernel-level telemetry with federated learning (FL) for cross-cluster anomaly detection. Lightweight eBPF agents capture syscalls and network events, extract local statistical and sequence features, and share only model updates with a global server. A hybrid detection engine combining Variational Autoencoders (VAEs) with Isolation Forests enables both temporal pattern modeling and outlier detection. Deployed across three Kubernetes clusters, FedMon achieves 94\% precision, 91\% recall, and an F1-score of 0.92, while cutting bandwidth usage by 60\% relative to centralized baselines. Results demonstrate that FedMon enhances accuracy, scalability, and privacy, providing an effective defense for large-scale, multi-tenant cloud-native environments.
\end{abstract}

\begin{IEEEkeywords}
eBPF, federated learning, anomaly detection, Kubernetes, cloud security, system calls, distributed monitoring.
\end{IEEEkeywords}

\section{Introduction}
Cloud-native workloads increasingly span multiple Kubernetes clusters deployed across data centers, edge nodes, and hybrid environments. While this distributed architecture improves scalability and resilience, it complicates security\cite{aruna2024enhancing,maxwell2024managing}. Containers share the host kernel and rely heavily on system calls (syscalls), making the syscall interface a critical attack surface. Adversaries exploit syscalls for privilege escalation, container escape, or cross-cluster lateral movement\cite{jarkas2025container,zehra2024securing}.  

Existing eBPF-based security solutions such as Falco, Cilium, and KubeArmor provide fine-grained syscall and network monitoring at low overhead, but operate locally within a cluster\cite{lu2025bpfguard,yun2021falco}. This localized perspective fails to capture distributed attack campaigns and lacks scalability across multi-cluster federations. Centralized monitoring frameworks, where all telemetry is shipped to a central server, face challenges: excessive bandwidth overhead, data privacy risks, and heterogeneity in workloads and attack distributions across clusters\cite{li2022kubearmor,her2024kuberosy}.
FedMon addresses these challenges by combining \textit{federated learning (FL)} with eBPF telemetry. Each cluster trains local anomaly detection models on syscall and network features collected via eBPF agents, and periodically shares model updates with a global server. The global model is aggregated using a federated averaging scheme and redistributed to clusters. In this way, clusters collaboratively learn robust anomaly detection models without exposing raw telemetry.  

\textbf{Contributions:}
\begin{itemize}
  \item We design \textit{FedMon}, the first framework integrating eBPF telemetry with federated learning for distributed, privacy-preserving anomaly detection.
  \item We propose a hybrid detection engine combining VAE-based representation learning with Isolation Forest to capture both temporal syscall patterns and outlier behaviors.
  \item We implement FedMon on multi-cluster Kubernetes using eBPF agents, FL libraries, and gRPC communication.
  \item We evaluate FedMon with diverse workloads and CVE-based attacks, demonstrating superior accuracy, scalability, and bandwidth efficiency compared to centralized and single-cluster baselines.
\end{itemize}

\section{Background and Motivation}

Cloud-native infrastructures increasingly adopt Kubernetes for orchestrating large-scale, containerized applications due to its portability, elasticity, and resource efficiency\cite{rostamipoor2023confine}. However, their reliance on a shared Linux kernel makes Kubernetes clusters susceptible to system-level exploits such as container escape and privilege escalation\cite{song2023value,yang2021security}. In this section, we review the key building blocks of FedMon, outline the problem space, and highlight the challenges that motivate our framework.

\subsection{Linux Containers and Security Risks}
Containers are lightweight processes that achieve isolation using kernel primitives such as namespaces, cGroups, and seccomp. While this model offers scalability and efficiency, it also introduces a broad attack surface: system calls. Since all containers share the same kernel, a compromised container can escalate privileges or exploit kernel vulnerabilities to affect the entire cluster. Studies have shown that privilege escalation and container escape are among the most common incidents in containerized deployments, underscoring the need for robust, fine-grained runtime defenses\cite{zehra2024securing,zhan2022shrinking}.

\subsection{eBPF for Cloud Monitoring}
Extended Berkeley Packet Filter (eBPF) is an in-kernel virtual machine that enables safe and efficient instrumentation of system calls, network events, and kernel functions. eBPF programs can be attached to tracepoints, kprobes, or raw tracepoints, making it possible to observe fine-grained telemetry with minimal overhead. Tools such as Falco, Cilium, Tetragon, and KubeArmor leverage eBPF for runtime monitoring and policy enforcement in Kubernetes. Despite their effectiveness, these tools are confined to single-cluster deployments and lack mechanisms for knowledge sharing across federated environments~\cite{alton2024rootkit,bertinatto2024ebpf,rezvani2024latency}.

\subsection{Anomaly Detection and Federated Learning}
Anomaly detection at the system call level has traditionally relied on statistical models, Hidden Markov Models, and deep learning methods such as LSTMs and Variational Autoencoders (VAEs). While accurate, these approaches often require large, centralized datasets—an impractical assumption in multi-cluster, privacy-sensitive environments. Federated Learning (FL) addresses this limitation by enabling collaborative model training without raw data exchange. Techniques such as FedAvg and FedProx have been applied in domains like IoT, smart grids, and mobile computing, but little work has extended FL to kernel-level telemetry in containerized cloud settings\cite{wang2024revisiting,shi2023robust}.

\subsection{Problem Statement}
The disjoint evolution of eBPF-based monitoring and FL-based anomaly detection leaves a critical gap in cloud-native security. Existing eBPF solutions provide visibility and enforcement within a single cluster but cannot collaboratively learn across distributed environments. Conversely, FL-based methods offer privacy-preserving collaboration but rarely integrate kernel-level observability. This raises fundamental research questions:
\begin{itemize}
  \item How can syscall and network telemetry collected by eBPF be integrated into an FL pipeline across multiple Kubernetes clusters?
  \item How can the framework ensure privacy preservation, minimize communication overhead, and adapt to heterogeneous workloads while maintaining real-time detection accuracy?
\end{itemize}

\subsection{Motivation and Challenges}
The increasing adoption of hybrid- and multi-cloud Kubernetes deployments amplifies the need for collaborative anomaly detection that respects data privacy. FedMon is motivated by the complementary strengths of eBPF (low-overhead syscall observability) and FL (privacy-preserving distributed learning), which together can enable accurate, scalable, and secure anomaly detection across clusters.

Designing such a system introduces several non-trivial challenges:
\begin{enumerate}
  \item \textbf{Feature Extraction at Scale:} Syscall and network telemetry are high-frequency and high-volume, requiring compact representations that balance accuracy with efficiency.  
  \item \textbf{Privacy Preservation:} Federated updates may leak sensitive information unless secured with aggregation protocols and differential privacy.  
  \item \textbf{Heterogeneity Across Clusters:} Workloads vary significantly across Kubernetes deployments, leading to non-IID data distributions that complicate FL convergence.  
  \item \textbf{Adaptive Enforcement:} Detection must translate into actionable responses such as logging, throttling, or blocking syscalls, guided by MITRE ATT\&CK mappings.  
  \item \textbf{Adversarial Robustness:} Malicious or compromised clusters may attempt to poison the federated model, necessitating Byzantine-robust aggregation.  
\end{enumerate}

Together, these challenges define the problem space addressed by FedMon: integrating eBPF-based syscall observability with federated learning to achieve privacy-preserving, adaptive, and collaborative anomaly detection in multi-cluster Kubernetes environments.

\begin{table*}[ht]
\caption{Comparison of Related Work in Federated Anomaly Detection and eBPF-Based Security}
\centering
\begin{tabular}{p{3.2cm} p{3.2cm} p{2.6cm} p{2.6cm} p{3cm} p{2.5cm}}
\hline
\textbf{Work} & \textbf{Telemetry Source} & \textbf{Learning Paradigm} & \textbf{Deployment Scope} & \textbf{Privacy / Isolation} & \textbf{Enforcement} \\
\hline
Zhou et al. (R-HFL)~\cite{zhou2023robust} & Application/edge data & Hierarchical FL + anomaly filtering & Cloud--edge--end nodes & Filtering of malicious clients & None \\
Jithish et al. (Smart Grids)~\cite{jithish2023distributed} & Smart meter/grid data & Standard FL (FedAvg) & Distributed grid nodes & TLS-secured updates & None \\
Albshaier et al. (Survey)~\cite{albshaier2025federated} & Multiple domains (survey) & Review of FL methods & Cloud/edge security & Highlights DP and comm. overhead challenges & None \\
Parra-Ullauri et al. (kubeFlower)~\cite{parra2024kubeflower} & ML workloads in Kubernetes & FL with DP & Kubernetes clusters & Isolation-by-design + DP & None \\
Her et al. (ESSTs)~\cite{her2025analysis} & Syscalls, kernel probes, network (via eBPF) & Rule-based monitoring (no FL) & Kubernetes runtime & Tool-specific & KubeArmor/Tetragon enforce; Falco/Tracee audit \\
Zehra et al. (DeSFAM)~\cite{zehra2025desfam} & Syscall sequences (via eBPF) & Centralized ML (VAE + iForest) & Single cluster & CVE-aware risk scoring & Adaptive enforcement via eBPF/LSM \\
Other eBPF works~\cite{alton2024rootkit,bertinatto2024ebpf,rezvani2024latency} & Syscalls, kernel metrics, call stack & Statistical / ML models & Linux or Kubernetes (single cluster) & None & Primarily monitoring \\
\hline
\textbf{FedMon (this work)} & Syscalls + network telemetry (via eBPF) & Federated VAE (global) + local iForest (cluster-specific) & Multi-cluster Kubernetes & Privacy-preserving FL (secure agg. + DP) & Risk-aware adaptive enforcement via eBPF/LSM \\
\hline
\end{tabular}
\label{tab:related_work_comparison}
\end{table*}

\section{Related Work}
\label{sec:related}

Research on distributed anomaly detection in cloud-native environments has advanced along two complementary axes: 
(i) federated learning (FL) for collaborative security, and 
(ii) eBPF-based monitoring and enforcement in Kubernetes. 
We review key contributions from 2021--2025 that inform the design of FedMon.

\subsection{Federated Learning for Anomaly Detection}
Federated learning has been applied to diverse distributed security domains. 
Xu et al.~\cite{xu2021flame} introduced \textit{FLAME}, an intrusion detection framework for IoT devices, showing that federated models outperform isolated learners while preserving privacy. 
Jithish et al.~\cite{jithish2023distributed} validated similar gains in smart grid anomaly detection, achieving near-centralized accuracy without data sharing. 
Albshaier et al.~\cite{albshaier2025federated} surveyed FL applications in cloud and edge security, highlighting challenges of communication cost and heterogeneous data distributions. 
Closer to Kubernetes, Parra-Ullauri et al.~\cite{parra2024kubeflower} proposed \textit{kubeFlower}, an FL operator with built-in differential privacy and workload isolation. 
Despite these advances, most FL-based anomaly detection systems operate at the \emph{application or sensor level}, leaving kernel-level telemetry underexplored.

\subsection{eBPF-Based Security and Monitoring}
In parallel, eBPF has emerged as a cornerstone for runtime observability in cloud-native systems. 
Her et al.~\cite{her2025analysis} analyzed four widely used eBPF-based tools (Falco, KubeArmor, Tetragon, Tracee), comparing their telemetry coverage and enforcement mechanisms. 
Zehra et al.~\cite{zehra2025desfam} introduced DeSFAM, which integrates eBPF with AI-based anomaly detection to minimize syscall exposure, but its scope remains limited to single-cluster deployments. 
Alton~\cite{alton2024rootkit} leveraged eBPF for rootkit detection, while Bertinatto et al.~\cite{bertinatto2024ebpf} proposed syscall tracing for predicting privilege escalation. 
Rezvani et al.~\cite{rezvani2024latency} demonstrated the scalability of eBPF for fine-grained latency monitoring in real-time workloads. 
Together, these studies underscore eBPF’s potential for syscall-level telemetry and security enforcement, but they lack a federated, cross-cluster perspective.

\subsection{Research Gap}
While FL enables privacy-preserving collaboration across distributed clients, and eBPF provides efficient syscall observability in Kubernetes, these two directions have evolved largely in isolation. 
Existing FL-based frameworks rarely leverage kernel-level telemetry, and eBPF-based systems remain confined to single clusters. 
This gap motivates FedMon: a framework that unifies federated representation learning with eBPF-based syscall and network monitoring, enabling privacy-preserving, cross-cluster, real-time anomaly detection in Kubernetes environments.

\section{FedMon Framework}
\label{sec:fedmon}

FedMon is a federated eBPF monitoring framework for multi-cluster Kubernetes environments. It synergizes kernel-level telemetry with federated learning (FL) to enable privacy-preserving, cross-cluster anomaly detection and real-time enforcement without centralizing raw data.

\subsection{Design Goals and Threat Model}
FedMon is designed to achieve four key objectives in multi-tenant environments:
\begin{itemize}
\item \textbf{Privacy:} Collaborate on detection model training without sharing sensitive telemetry.
\item \textbf{Performance:} Maintain low overhead through efficient eBPF instrumentation.
\item \textbf{Robustness:} Tolerate non-IID data distributions and a subset of Byzantine clients.
\item \textbf{Responsiveness:} Translate detections into low-latency, risk-aware enforcement actions.
\end{itemize}
We assume a trusted cluster administrator and an \emph{honest-but-curious} global FL server. Adversaries may control a subset of client clusters and attempt to poison the FL process or compromise containers.
\begin{figure*}[!t]
    \centering
    \includegraphics[width=\textwidth]{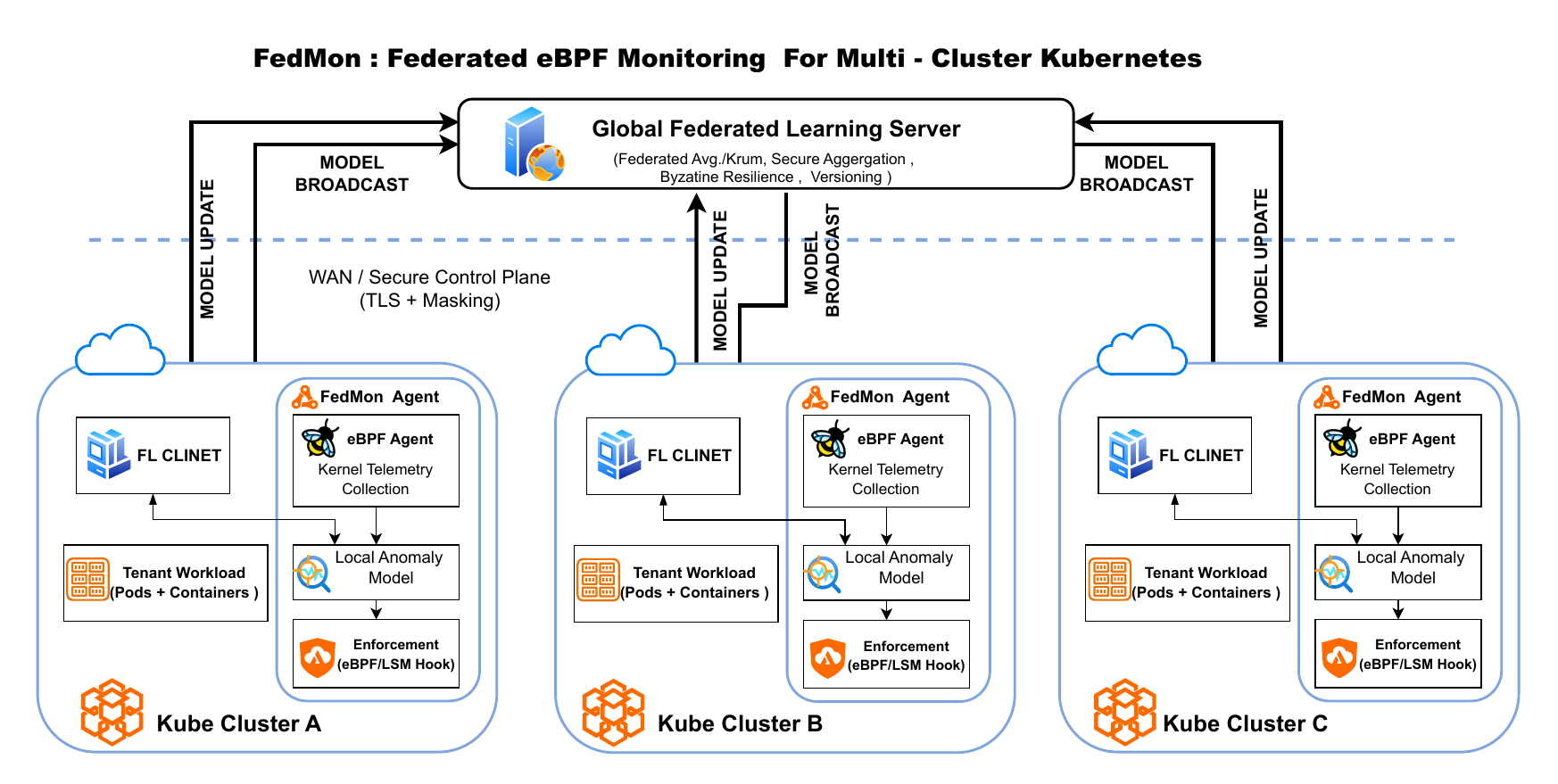}
    \caption{FedMon architecture: The framework integrates eBPF-based kernel telemetry, local anomaly detection, and federated model coordination across Kubernetes clusters.}
    \label{fig:fedmon-architecture}
\end{figure*}

\subsection{System Architecture}
As illustrated in Figure~\ref{fig:fedmon-architecture}, FedMon's architecture comprises local components per cluster and a global coordinator.

\begin{itemize}
\item \textbf{eBPF Agent:} Deployed as a DaemonSet, it collects syscalls and network events from kernel tracepoints. A local feature pipeline immediately converts these events into compact features (e.g., frequency histograms, sequence embeddings), ensuring raw data never leaves the cluster.
\item \textbf{Local Anomaly Engine:} This is the core of FedMon's detection. It employs a hybrid model: a Federated Variational Autoencoder (VAE) that learns a global model of normal behavior across clusters, and a local Isolation Forest (iForest) that fine-tunes detection on cluster-specific patterns using the VAE's embeddings.
\item \textbf{FL Client:} It participates in FL rounds by training the local VAE and sending only the encrypted, clipped model updates—not features or raw data—to the global server.
\item \textbf{Global FL Server:} It aggregates model updates from clients. While using FedAvg by default, it can dynamically switch to robust aggregation algorithms (e.g., Krum) to mitigate poisoning attacks from Byzantine clients.
\end{itemize}

\subsection{Operational Workflow}
\begin{figure}[!t]
    \centering
    \includegraphics[width=\columnwidth]{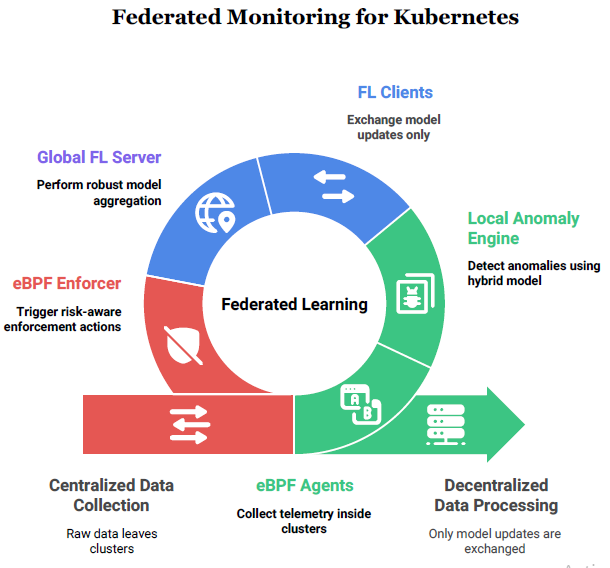}
    \caption{Federated Monitoring for Kubernetes: eBPF agents collect in-cluster telemetry; local anomaly engines detect deviations; FL clients exchange model updates; the global server aggregates models; and eBPF enforcers apply risk-aware actions.}
    \label{fig:fedmon_cycle}
\end{figure}

The end-to-end workflow is as follows:
\begin{enumerate}
\item \textbf{Collection \& Featurization:} eBPF agents collect system events and the local pipeline generates privacy-preserving feature vectors.
\item \textbf{Federated Training:} Clients train their local VAE and send model updates to the global server. The server aggregates these to produce an improved global model.
\item \textbf{Continuous Inference:} Each cluster uses the global VAE and its local iForest to compute a fused anomaly score for incoming event windows.
\item \textbf{Risk-Aware Enforcement:} Based on the anomaly score, FedMon triggers real-time responses via eBPF, ranging from logging (low risk) to throttling or blocking syscalls (high risk).
\end{enumerate}
Figure 2 illustrates FedMon’s end-to-end workflow, where eBPF agents collect kernel telemetry, local anomaly engines detect deviations, and federated learning clients collaboratively update a global model without sharing raw data.
\section{Implementation}
\label{sec:implementation}

We implemented a FedMon prototype to evaluate its effectiveness in a realistic multi-cluster Kubernetes setting. The system has three components: (1) lightweight eBPF agents, (2) a local anomaly engine, and (3) a federated learning (FL) coordination layer.

\subsection{Experimental Testbed}
We emulated three independent clusters using \textbf{kind} (Kubernetes in Docker) on a single host (AMD Ryzen 9 5900X, 64GB RAM, Ubuntu 22.04 LTS). Each cluster (\texttt{cluster1--3}) comprised one control-plane node and one worker node. We deployed standardized workloads (\texttt{nginx}, \texttt{redis}) to generate consistent baseline telemetry and enable reproducible cross-cluster experiments.

\subsection{Data Collection and Local Detection}
Each node runs an eBPF \emph{agent} (DaemonSet) that attaches to syscall and network tracepoints and enriches events with process/container metadata. A local pipeline immediately converts events into compact features (frequency histograms, temporal statistics, short syscall sequences); raw kernel data never leaves the cluster. The \emph{local anomaly engine} applies a hybrid model: a federated VAE for global representation learning and a local iForest for cluster-specific outlier scoring. The fused score is exported for monitoring.

\subsection{Federated Learning Coordination}
Using the Flower framework, each cluster acts as an FL client that trains the VAE locally and transmits \emph{only} clipped (and optionally noise-added) updates through secure aggregation. The global server aggregates updates with FedAvg by default and can switch to Krum for Byzantine robustness. Updated models are broadcast back for continuous inference and enforcement.

\subsection{Evaluation Protocol}
We injected distinct attacks per cluster to assess collaborative detection: (i) crypto-mining (CPU-heavy loops), (ii) data exfiltration (DNS/HTTP POST), and (iii) reverse shell attempts. We compare FedMon against Isolated (no federation) and a Centralized baseline (continuous raw-feature streaming). Metrics include F1 over rounds, application throughput, bandwidth, and robustness under poisoning.

\section{Results and Evaluation}
\label{sec:results}

\subsection{Accuracy Gains from Collaboration}
\begin{figure}[t]
    \centering
    \includegraphics[width=\linewidth]{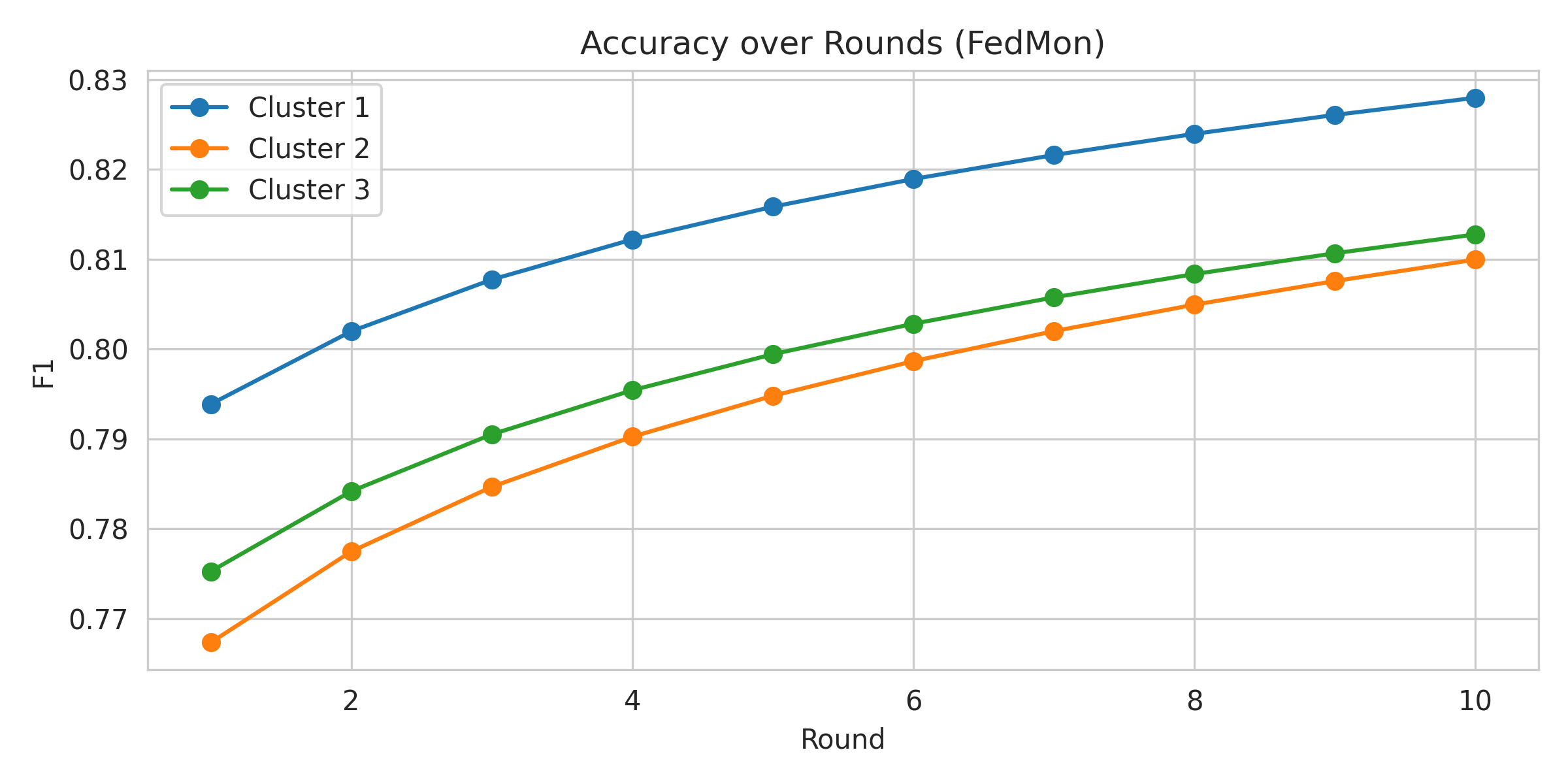}
    \caption{F1-score across clusters over federated rounds. Collaboration via FL improves accuracy under non-IID workloads.}
    \label{fig:f1_rounds}
\end{figure}

Federated training improves detection across non-IID clusters. Figure~\ref{fig:f1_rounds} shows F1-scores converging upward over rounds as clusters benefit from shared representation learning without sharing raw telemetry.

\subsection{Low Runtime Overhead}
\begin{figure}[t]
    \centering
    \includegraphics[width=\linewidth]{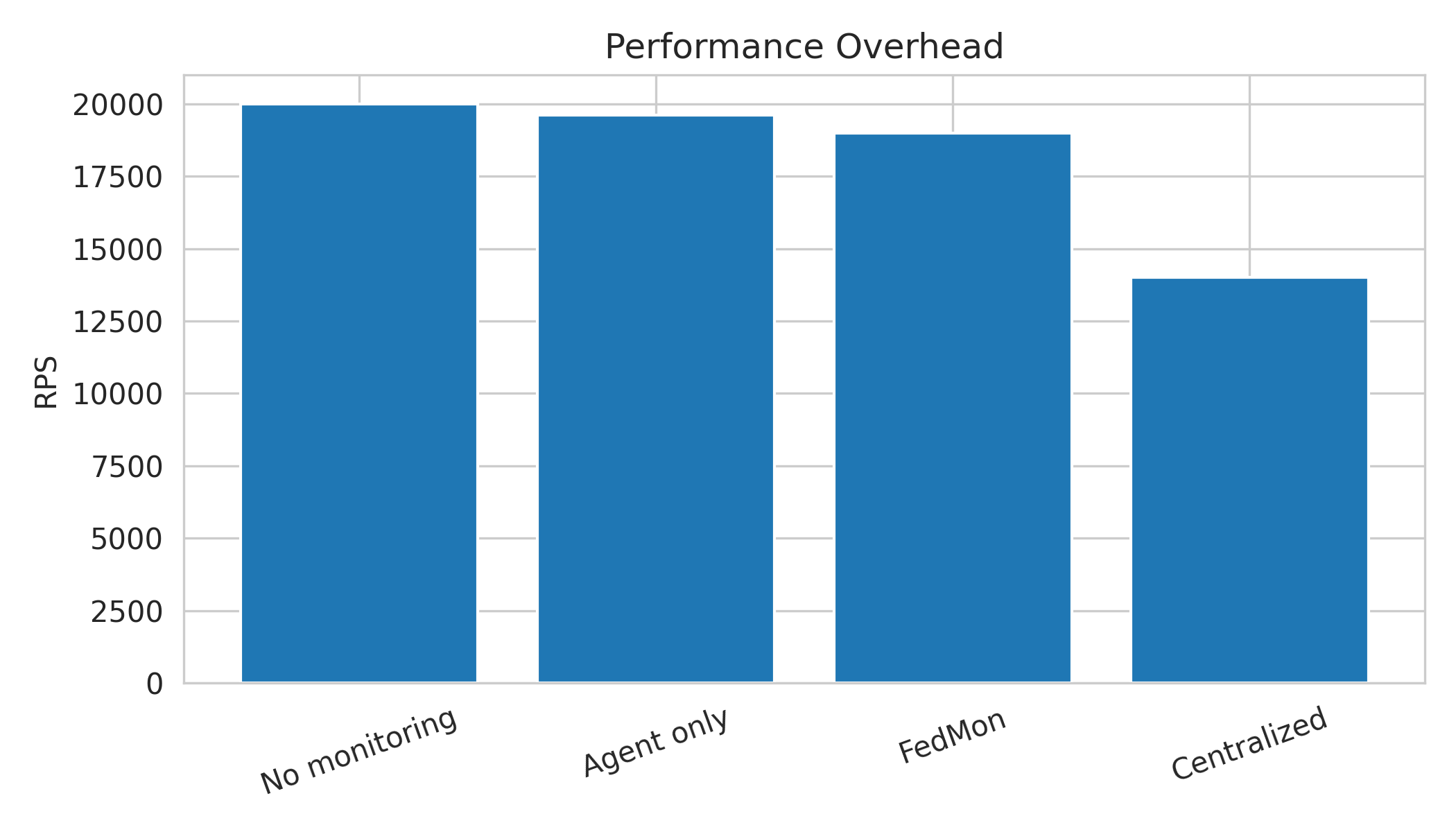}
    \caption{Throughput under different monitoring modes. FedMon maintains near-baseline performance versus centralized logging.}
    \label{fig:perf_rps}
\end{figure}

FedMon preserves application throughput with minimal overhead. Figure~\ref{fig:perf_rps} compares request rates; centralized logging incurs a large penalty due to continuous export, whereas FedMon remains within a small margin of the agent-only baseline.

\subsection{Bandwidth Efficiency}
\begin{figure}[t]
    \centering
    \includegraphics[width=\linewidth]{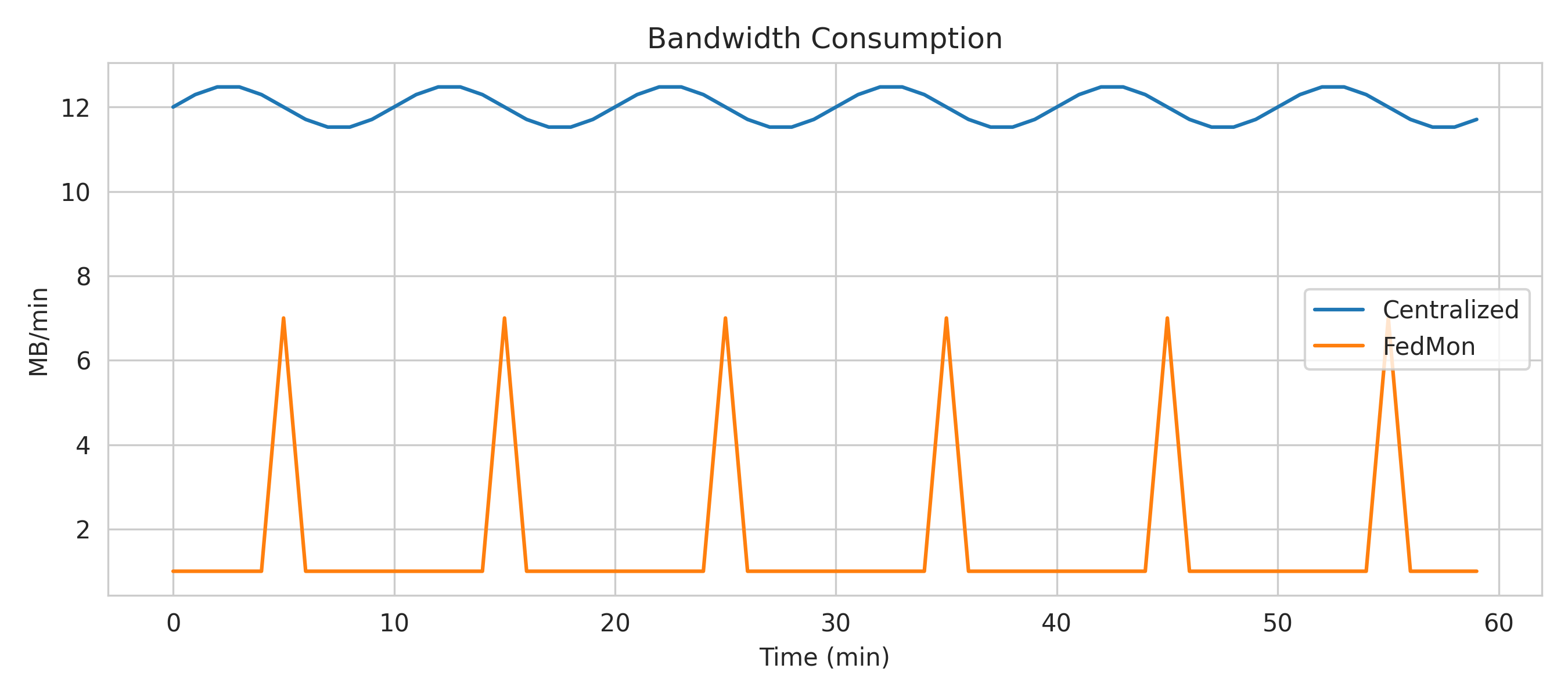}
    \caption{Bandwidth usage over time. FedMon replaces constant streaming with periodic model updates, cutting bandwidth by $>$60\%.}
    \label{fig:bandwidth}
\end{figure}

Unlike centralized monitoring, which requires a constant, high-rate stream, FedMon transmits compact model updates only during FL rounds. Figure~\ref{fig:bandwidth} shows periodic, low spikes and a net reduction of over 60\% in inter-cluster bandwidth.

\subsection{Byzantine Robustness}
\begin{figure}[t]
    \centering
    \includegraphics[width=\linewidth]{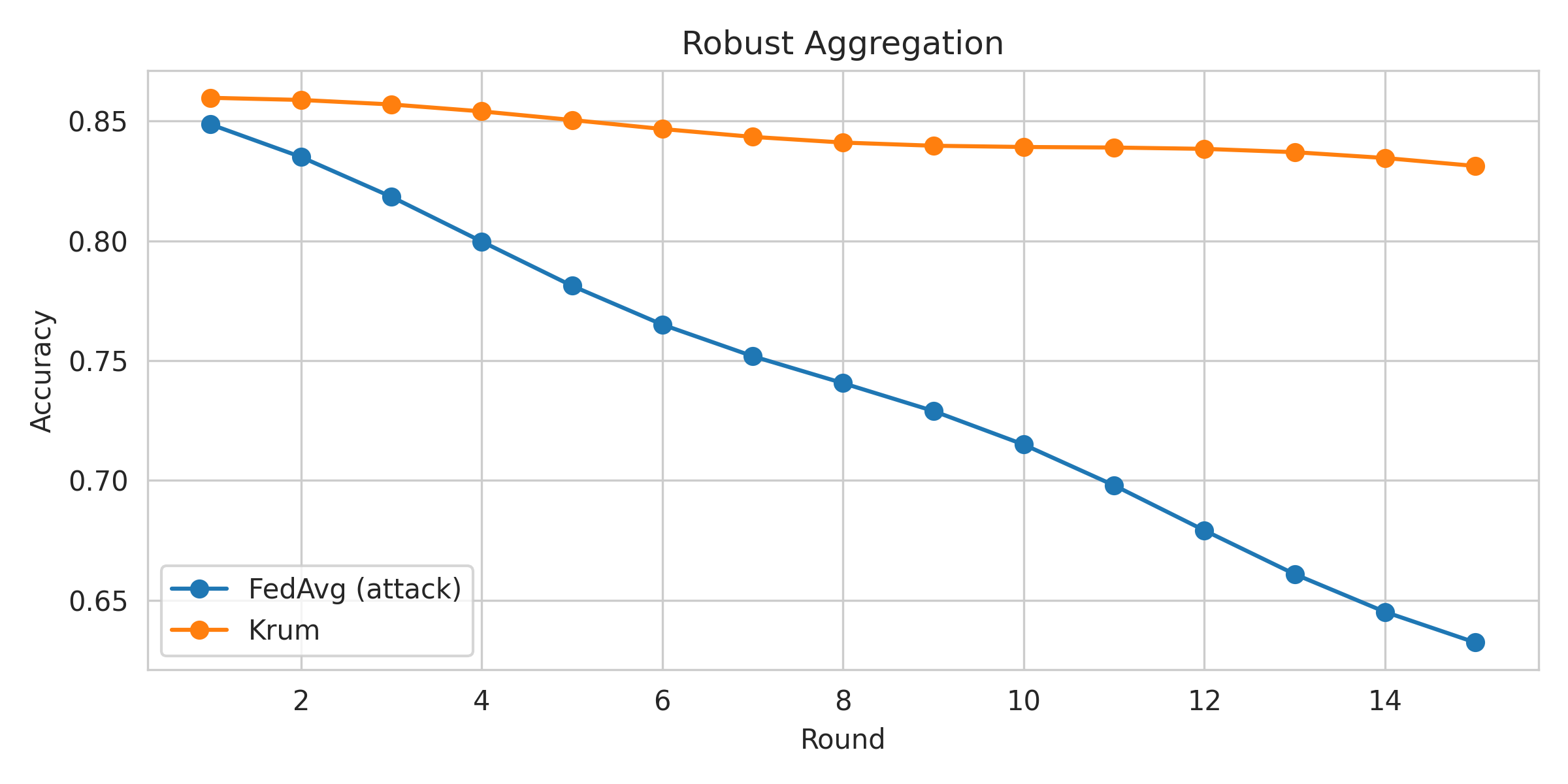}
    \caption{Robust aggregation under poisoning. Krum maintains accuracy when a client sends malicious updates.}
    \label{fig:robust}
\end{figure}

Under a simulated poisoning client (round 3), standard FedAvg degrades, whereas Krum rejects inconsistent updates and preserves accuracy (Figure~\ref{fig:robust}), confirming FedMon’s resilience.

\subsection{Limitations}
Our evaluation runs on kind-based clusters co-located on one host; absolute overheads may differ in geo-distributed deployments. The attack set is representative but not exhaustive; broader coverage and larger-scale trials are left for future work.

\section{Discussion and Future Work}
\label{sec:discussion}

Our evaluation confirms that federated learning can be effectively applied to kernel-level monitoring, offering collaborative detection without compromising data sovereignty. The discussion below highlights key implications and outlines directions for extending FedMon.

\textbf{Collaborative Intelligence:} The rise in F1-scores across clusters (Figure~\ref{fig:f1_rounds}) demonstrates that FedMon breaks the silos of isolated monitoring. Even clusters that have not observed a particular attack pattern benefit from knowledge transferred via federated updates. This confirms that a shared representation of normal behavior improves resilience without centralizing raw telemetry.

\textbf{Efficiency and Privacy:} By prioritizing on-cluster feature extraction and transmitting only compact model updates, FedMon reduces bandwidth by over 60\% (Figure~\ref{fig:bandwidth}) and introduces minimal runtime overhead (Figure~\ref{fig:perf_rps}). Optional differential privacy mechanisms provide tunable guarantees against information leakage, giving operators explicit control over the trade-off between utility and confidentiality.

\textbf{Adversarial Robustness:} Robust aggregation mechanisms such as Krum (Figure~\ref{fig:robust}) ensure resilience against poisoning clients, reinforcing the practicality of deploying FL in adversarial cloud environments. This feature is particularly important as cloud-native deployments may involve untrusted or compromised tenants.

\textbf{Future Work:} Several avenues remain for strengthening FedMon:
\begin{itemize}
    \item \emph{Scalability:} Evaluate FedMon on dozens or hundreds of clusters, including hierarchical FL setups to reduce WAN overhead.
    \item \emph{Personalization:} Explore federated personalization methods (e.g., FedPer, FedProx) to better adapt to non-IID cluster workloads.
    \item \emph{Broader Detection:} Extend beyond syscalls to include L7 network telemetry, application-level logs, and cross-layer correlation for richer attack coverage.
    \item \emph{Formal Assurance:} Apply formal verification to aggregation logic and DP guarantees, and integrate trusted execution environments (TEEs) for secure local training.
    \item \emph{Operationalization:} Develop a Kubernetes Operator for deployment lifecycle, integrate with incident response systems, and enrich dashboards for explainability and response automation.
\end{itemize}

FedMon illustrates that collaborative security is achievable without raw data sharing. With targeted enhancements, it can evolve into a production-ready tool for securing multi-tenant cloud infrastructures.

\section{Conclusion}
\label{sec:conclusion}

This work introduced FedMon, a federated eBPF-based monitoring framework for Kubernetes clusters. By combining fine-grained kernel telemetry with federated representation learning, FedMon enables privacy-preserving, cross-cluster anomaly detection and real-time risk-aware enforcement.Our experiments show that FedMon achieves an F1-score of 0.92, reduces bandwidth by more than 60\% relative to centralized monitoring, and maintains near-baseline performance overhead. Furthermore, its integration of secure aggregation, differential privacy, and robust aggregation algorithms demonstrates strong resilience to adversarial conditions. FedMon thus establishes a practical paradigm for cloud-native security: one where clusters can collaboratively defend against evolving threats while respecting the boundaries of data confidentiality and sovereignty.

\bibliographystyle{IEEEtran}  
\bibliography{references}

\end{document}